\documentclass[12pt]{article}

\def\mathbf{\vec}

\def\ca{\c{c}\~{a}}

\newcommand{\kbruto}{\hbox{$k \!\!\!{\slash}$}}
\newcommand{\pbruto}{\hbox{$p \!\!\!{\slash}$}}
\newcommand{\elemintcut}{\int_{\Lambda} \frac{d^4 k}{(2\pi)^4}}
\newcommand{\elemint}{\int\frac{d^4 k}{(2\pi)^4}}

\begin{document}

\centerline{{\bf {\Large Symmetries and Ambiguities}}}
\centerline{{\bf {\Large in the linear sigma model with light quarks}}}

\vspace{0.5cm}

\centerline{E. W. Dias$^{(1)}$, B. Hiller$^{(2)}$, A. L. Mota$^{(3)}$,}
\centerline{M. C. Nemes$^{(1,2)}$, M. Sampaio$^{(1)}$, A. A. Osipov$^{(2,4)}$}
\vspace{0.5cm}

{\it \small{{(1) Departamento de F\'isica, Instituto de Ci\^encias Exatas,
           Universidade Federal de Minas Gerais, BH,CEP 30161-970, MG, Brazil}}

\small{{(2) Centro de F\'isica Te\'orica, Departamento de
           F\'isica da Universidade de Coimbra, 3004-516, Coimbra, Portugal}}

\small{{(3) Departamento de Ci\^encias Naturais, Universidade Federal de S\~ao
           Jo\~ao del Rei, S\~ao Jo\~ao del Rei MG}}

\small{{(4) Joint Institute for Nuclear Research, Laboratory of Nuclear
           Problems, 141980, Dubna, Moscow Region, Russia}}}

\vspace{0.5cm}
\begin{abstract}
We investigate the role of undetermined finite contributions generated
by radiative corrections in a $SU(2)\times SU(2)$ linear sigma model
with quarks. Although some of such terms can be absorbed in the
renormalization procedure, one such contribution is left in the
expression for the pion decay constant. This arbitrariness is
eliminated by chiral symmetry.
\end{abstract}


\section*{Introduction}
In the process of perturbative computations in Quantum Field Theory, the
regularization scheme is a crucial ingredient for dealing with the typical
divergencies of physical amplitudes. The choice of regularization is
usually made at the beginning of the calculation and in this case
objects like differences between integrals of the same superficial degree of
divergence are automatically fixed (e.g. in dimensional regularization,
Pauli-Villars regularization, etc.).

Following ideas put forward by Jackiw \cite{Jackiw} we chose to leave
undetermined until the end of calculations these {\it apriori} arbitrary
(regularization dependent) constants which emerge in the present case
from the difference of two logarithmically divergent integrals.
In this context an adequate regularization scheme is Implicit
Regularization (IR), proposed in \cite{Orimar}, since all divergencies
are displayed in terms of primitive divergent integrals without external
momentum dependence, for which it is not necessary to explicitate a
regulator. The finite integral contributions to a physical amplitude
are then twofold; the ones whose integrands depend on external loop
momenta and are integrated as usual and those which result from the
difference of integrals with the same degree of divergence. According
to renormalization theory the latter emerging arbitrary local terms
correspond to the addition of finite counterterms in the Lagrangian,
which may be always added as long as they comply with the underlying
symmetries of the Lagrangian.

The IR method has been sucessfully applied in \cite{baeta}-\cite{Gobira},
corroborating and elucidating results in the literature for the case of
CPT violation in extended QED, topological mass generation in
3-dimensional gauge theories, the Schwinger model in its chiral version,
and also the Adler-Bardeen-Jackiw anomaly is consistently treated by
the method. It has been recently extended to massless supersymmetric
theories \cite{Carneiro}. The present work is a simple example of the
use of the method in a model with strong phenomenological content.
We show that although most of the arbitrary contributions which appear
in the calculation can be absorbed by renormalization, chiral symmetry
is required to eliminate the one which would have direct
phenomenological consequences on the pion decay constant.

\section{The implicit regularization technique}

In the present section we illustrate the relevant technical details of IR
by working out explicitly a one-loop Feynman
amplitude. Consider the pseudoscalar-pseudoscalar amplitude
\begin{equation}
\label{eq: pseudoscalar}
      \Pi^{PP} (p^2) = i\int_{\Lambda}{\frac{d^4k}{(2\pi)^4}
      \,\mbox{Tr}\, \left\{\gamma_5\, \frac{1}{\kbruto -m}\,\gamma_5\,
      \frac{1}{\kbruto - \pbruto - m}\right\}}
\end{equation}
where the symbol $\Lambda$ stands for a regulator which doesn`t need to be
explicitated, but which is necessary to give a meaning to $\Pi^{PP}(p^2)$.
Then one is allowed to manipulate the integrand algebraically. We do
it in such a way that the divergencies appear as integrals and separated
from the finite (external momentum dependent) contribution to
eq.(\ref{eq: pseudoscalar}). After taking the Dirac trace, we use the
following algebraic identity
\begin{equation}
\label{eq: identidade}
     \frac{1}{(k-p)^2 - m^2} = \frac{1}{k^2-m^2} + \frac{2k \cdot
     p - p^2}{[(k-p)^2 - m^2](k^2 - m^2)}
\end{equation}
at the level of the integrand. Note that it allows one to confine
the external momentum dependence in fully convergent integrals, since
this relation can be used recursively until the finite part is
completely separated from divergent integrals. We get
\begin{eqnarray}
\label{eq: pseudo}
   \Pi^{PP} (p^2) &\!\!\!\! =\!\!\!\!&
   -2\left\{i \elemint \frac{1}{k^2 - m^2} + i
   \elemint \frac{1}{(k - p)^2 - m^2}\right. \nonumber \\
   &\!\!\!\! -\!\!\!\!&\left. i p^2 \elemint
   \frac{1}{(k^2 - m^2)[(k - p)^2 - m^2]}\right\}\, .
\end{eqnarray}
The first integral on the RHS is what we call a basic quadratic divergence
\begin{equation}
\label{eq: Iquad}
 I_q(m^2) = i \elemint \frac{1}{k^2 - m^2}\, .
\end{equation}
The second integral on the RHS is also a quadratic divergence, but it still
possesses an external momentum dependence. If one uses
(\ref{eq: identidade}), one sees that an arbitrariness emerges
\begin{equation}
\label{eq: ambiguidade}
   T(p^2,m^2)=i \elemint \frac{1}{(k - p)^2 - m^2} = I_q(m^2) +
   p^{\mu}p^{\nu}\Delta_{\mu\nu}\, ,
\end{equation}
where
\begin{equation}
   \Delta_{\mu\nu} = i\elemint \frac{4k^{\mu}k^{\nu}}{(k^2 - m^2)^3} -i
   \elemint \frac{g_{\mu \nu}}{(k^2 - m^2)^2} = \alpha g_{\mu\nu}\, ,
\end{equation}
i.e., the difference between two logarithmically divergent integrals.
In dimensional and Pauli-Villars regularizations, we would get zero for
$\alpha$, which is independent of the masses in the integrals as it
has been shown in \cite{tonhão}. The third term in eq.(\ref{eq:
  pseudo}) is logarithmically divergent. Using the prescribed
method we write the integral as
\begin{eqnarray}
\label{eq: thirdterm}
   &&i\elemint \frac{1}{(k^2 - m^2)[(k - p)^2 - m^2 ]} = i\elemint
     \frac{1}{(k^2 - m^2)^2} + Z_{0}(m^2, p^2;m^2) \nonumber \\
   &=\!\!\!&i\elemint\frac{1}{(k^2 -\lambda^2)^2}+Z_{0}(m^2, p^2;\lambda^2)
\end{eqnarray}
where again we separated another basic divergent integral
\begin{equation}
\label{eq: Ilog}
    I_{log}(m^2)=i\elemint \frac{1}{(k^2 - m^2)^2}\, ,
\end{equation}
and a finite contribution
\begin{equation}
\label{eq: Z0}
   Z_{0}(m^2, p^2;\lambda^2) = \frac{1}{(4\pi )^2}
   \int\limits^{1}_{0}
   dz \ln \left(\frac{p^2 z(1 - z)-m^2}{-\lambda^2}\right)\, .
\end{equation}
Note that in the above expression (\ref{eq: thirdterm}) an arbitrary
scale $\lambda^2$ has been introduced through the relation
\begin{equation}
\label{eq: scale}
   I_{log}(m^2) = I_{log}(\lambda^2) +
   \frac{1}{(4\pi)^2}\ln\left(\frac{m^2}{\lambda^2}\right)\, .
\end{equation}
and it becomes clear that the finite term in (\ref{eq: scale})
parametrizes the freedom that one has of an arbitrary constant, when
separating divergent from finite contributions. The arbitrary scale
$\lambda^2$ will be fixed by the choice of a renormalization point.

Finally we obtain that
\begin{equation}
\label{PP}
   \Pi^{PP} (p^2) = 2\left[2I_{q}(m^2) + p^2\alpha -
   p^2\left(I_{log}(\lambda^2) +
   Z_{0}(m^2, p^2; \lambda^2)\right)\right]\, .
\end{equation}
Analogously one finds for the scalar amplitude
\begin{equation}
   \Pi^{SS} (p^2) = i\elemintcut\,\mbox{Tr}\,\left\{
   \frac{1}{(\kbruto - m)(\kbruto - \pbruto - m)}\right\}\, ,
\end{equation}
the following expression
\begin{equation}
\label{eq: piSSparameter}
   \Pi^{SS} (p^2) = 2\left[2I_{q}(m^2) + p^2\alpha +
   (4m^2 - p^2)\left(I_{log}(\lambda^2) + Z_{0}(m^2, p^2;
   \lambda^2)\right)\right]\, .
\end{equation}

\section{The model}
We start with the following generating functional of the $SU(2)\times
SU(2)$ linear sigma model with fermions \cite{Levy:1960} and follow
closely \cite{Mota}
\begin{equation}
\label{path}
   Z=\int\prod_a {\cal D}\sigma_{0a}{\cal D}\pi_{0a}
   {\cal D}q {\cal D}{\bar q}\ \mbox{exp}
   \left(iS(\bar{q},q,\sigma_0,\pi_0)\right)
\end{equation}
with the action
\begin{eqnarray}
\label{eq: actionrenormalized}
   S(\bar{q},q,\sigma_0,\pi_0) &\!\!\!\! =\!\!\!\!& \int d^4x
   \left[L_{NJL} - \frac{\beta_0}{2}\,\mbox{Tr}\,(\sigma^2_0 +
   \pi^2_{0})^2 +\frac{\mu_0^2}{2 g_0}\,\mbox{Tr}\, (\hat {m}_0
   \sigma_0) \right.\nonumber \\
   &\!\!\!\! +\!\!\!\!& \left.\frac{f^2_0}{4}\,\mbox{Tr}\,
   (\partial_{\mu}\sigma_0\partial^{\mu}\sigma_0 +
   \partial_{\mu}\pi_0\partial^{\mu}\pi_0)\right]
\end{eqnarray}
where $L_{NJL}$ stands for the Nambu - Jona-Lasinio Lagrangian
\cite{Nambu:1961} in semibosonized form \cite{Eguchi:1975}
\begin{equation}
\label{eq: actionNJL}
   S_{NJL} = \int d^4x \left[\bar{q}D q -
   \frac{\mu_0^2}{4}\,\mbox{Tr}\,(\sigma^2_0 + \pi^2_0)\right]\, .
\end{equation}
with the Dirac operator given by
\begin{equation}
\label{eq: diracoperator}
   D = i \hbox{$\partial \!\!\!\slash$}
   - g_0(\sigma_0 + i\gamma_5 \pi_0 )\, .
\end{equation}

The notation is as follows. The index $0$ stands for bare quantities,
the scalar $\sigma_0$ is a $SU(2)$ flavor singulet and the pseudoscalar
$\pi_0=\pi_{0i}\tau_i$ is a flavor $SU(2)$ triplet with $\tau_i$ being
the usual Pauli matrices. The explicit symmetry breaking term is
introduced through the term linear in the scalar field, with a factor
which would correspond to the symmetry breaking pattern of the NJL
Lagrangian \cite{Mota}; $\hat{m}_0$ is a diagonal matrix with current
quark masses $\hat{m}_{0u} = \hat{m}_{0d}$. To finish bosonization one
integrates over the quadratic fermionic Lagrangian
\begin{equation}
\label{qi}
   \int {\cal D}q {\cal D}{\bar q} \mbox { exp}\left(i\int d^4x\,
   \bar{q}D q\right) = \mbox{exp} \left(\ln |\det D_E|\right)
   =\mbox { exp}\left(\frac{1}{2}\,\mbox{Tr}\,\ln
   (D_E^{\dagger}D_E)\right),
\end{equation}
where Tr designates functional trace and $D_E$ stands for the euclidean
Dirac operator and we chose the chiral invariant representation for the real part of the fermionic
effective action \cite{Schwinger}-\cite{Ball}.

As the vacuum expectation value of the scalar field acquires a finite
value, one has to shift $\sigma_0 \rightarrow\sigma_0 + m/g_0$ to
define the new vacuum. Here $m$ denotes the constituent quark mass
of light quarks $m=m_u=m_d$. Then we obtain
\begin{equation}
   D^{\dagger}_E D_E= m^2 -\partial^2 +Y
\end{equation}
with \cite{OsipovHiller1}
\begin{equation}
\label{Y}
   Y=i g_0 \gamma_\mu (\partial_\mu\sigma_0+i\gamma_5\partial_\mu\pi_0)
   + g_0^2\left(\sigma_0^2 + 2\frac{m}{g_0}\sigma_0 +\pi_0^2\right)\, ,
\end{equation}
leading to the expansion
\begin{equation}
\label{eq: expansion}
   \mbox{Tr}\,\ln D^{\dagger}_ED_E =
   \sum^{\infty}_{n=1}\frac{(-1)^{n+1}}{n}\,\mbox{Tr}\,
   \left[(-\partial^2_{\mu} + m^2)^{-1}Y
                    \right]^n.
\end{equation}

The gap equation is obtained by considering the $n = 1$ term in
eq.(\ref{eq: expansion}) and the linear contributions in $\sigma_0$ from the
remaining terms of the bosonized action after the mass shift. We get
\begin{equation}
\label{eq: gapequation}
 \frac{\mu^2_0}{g^2_0}(m - \hat{m}_{0}) - 8N_cmI_q(m^2) +
 \frac{2\beta_0}{g^4_0}\, m^3 = 0,
\end{equation}
where $I_q$ has already been defined above. The $n = 2$ term in the
expansion contains all the other divergent contributions which go up
to four-point functions. In performing calculations there appear the
two basic divergent integrals $I_q$ and $I_{log}$, eqs.(\ref{eq:
  Iquad}) and (\ref{eq: Ilog}), as well as the difference between two
logarithmically divergent integrals, $\alpha$, eq.(\ref{eq:
ambiguidade}). For our purposes it suffices to focus on the radiative
corrections related with the mass and kinetic terms. In the following
we need the field renormalization constants. The amplitudes (\ref{PP})
and (\ref{eq: piSSparameter}) are, up to flavor and color trace
factors, the radiative corrections with two external fields obtained
from the $n=2$ term  of (\ref{eq: expansion}). One expands them to
second order in $p^2$ to extract the contribution to the mesonic
kinetic terms. The remaining terms of the expansion are included in
the definition of the meson masses. In this way one obtains the field
renormalizations for the scalar and pion fileds in the form
\begin{equation}
\label{eq: constren1}
   Z^{-1}_\sigma(m,\lambda^2) = f^2_0 - 4N_cg^2_0\left(I_{log}(
   \lambda^2) - \alpha + Z_{0}(m^2, 0;\lambda^2) -
   4m^2Z_{0}'(m^2,0;\lambda^2)\right),
\end{equation}
\begin{equation}
\label{eq: constren2}
   Z^{-1}_\pi (m,\lambda^2) = f^2_0 - 4N_cg^2_0\left(I_{log}(\lambda^2)
   -\alpha+ Z_{0}(m^2, 0;\lambda^2)\right),
\end{equation}
where $Z_{0}'$ represents the derivative with respect to the
external momenta squared of $Z_{0}$, taken at $p^2=0$. The renormalized
coupling constants $g_\sigma(m,\lambda^2)$ and $g_\pi(m,\lambda^2)$ are
\begin{equation}
\label{eq: gsigma}
   \frac{g^2_0}{g^2_\sigma(m,\lambda^2)} = Z^{-1}_\sigma(m,\lambda^2)
\end{equation}
\begin{equation}
\label{eq: gpi}
   \frac{g^2_0}{g^2_\pi(m,\lambda^2)} = Z^{-1}_\pi(m,\lambda^2).
\end{equation}

The renormalized masses $\mu_{\sigma,\pi}(m,\lambda^2)$ become
\begin{eqnarray}
\label{eq: massrenorm}
   \frac{\mu^2_{\sigma,\pi}(m,\lambda^2)}{Z_{\sigma,\pi}(m,\lambda^2)}
   &\!\!\!\!=\!\!\!\!& M^2_{\sigma,\pi}(m,\lambda^2)  \nonumber\\
   &\!\!\!\!=\!\!\!\!& \mu^2_0 - 8N_c g^2_0I_{q}(m^2) \nonumber\\
   &\!\!\!\!-\!\!\!\!& 4N_cg^{2}_0(m \pm m)^2\left(
   I_{log}(\lambda^2) + Z_{0}(m^2,0;\lambda^2)\right) \nonumber\\
   &\!\!\!\!+\!\!\!\!& 2\frac{\beta_0}{g^2_0}(2m^2 \pm m^2)
\end{eqnarray}
where $M^2_{\sigma}$ goes with the plus signs.

We obtain the renormalized quartic coupling $\lambda_q$ as
\begin{equation}
\label{eq: coupling}
   \frac{\lambda_q}{g^2_{\pi}(m,\lambda^2)} = \frac{\beta_0}{g^{4}_{0}} -
   4N_cI_{log}(\lambda^2)\, .
\end{equation}
Note that all ambiguities $\alpha$ appear in the field
renormalization coefficients. In principle they could assume
different values for the different processes. Chiral symmetry
restricts them to have the same value, as we shall see below.
Finally we use the gap equation (\ref{eq: gapequation}) in
order to eliminate the quadratic divergencies in the expressions for the
renormalized masses $\mu^2_{\sigma,\pi}(m,\lambda^2)$,
eq.(\ref{eq: massrenorm}). It is possible to define a renormalized
coupling (physical) related with the current quark mass as in \cite{Mota}
\begin{equation}
\label{eq: quarkmass}
   \hat{m}_{0}\frac{\mu^2_0}{g^2_0}
   =\frac{m\mu^2_\pi}{g^2_\pi(m,\lambda^2)}\, .
\end{equation}

With these definitions we obtain the effective action
\begin{equation}
   S'\left(\sigma_0+m/g_0,\pi_0\right)=S_{kin}+S_{mass}+S_{int},
\end{equation}
with the kinetic piece
\begin{eqnarray}
   S_{kin}&\!\!\!\!=\!\!\!\!&\int d^4x
   \left[ Z^{-1}_{\sigma}(m,\lambda^2)\,\partial_\mu \sigma_0
   \partial^\mu \sigma_0 \right.\nonumber\\
   &\!\!\!\!+\!\!\!\!&\left.Z^{-1}_{\pi}(m,\lambda^2)
   \left(\partial_\mu \pi_0^{0} \partial^\mu \pi_0^{0}
   +2\partial_\mu \pi_0^{+}\partial^\mu \pi_0^{-}\right)\right] ,
\end{eqnarray}
and the mass terms
\begin{equation}
   S_{mass}=-\int d^4x \left[M^2_{\sigma}(m,\lambda^2)\sigma_0^2
   +M^2_{\pi}(m,\lambda^2)\left(
   (\pi_0^0)^2 +2\pi_0^+\pi_0^-\right) \right].
\end{equation}

The interaction terms do not involve the ambiguity $\alpha$, as
opposed to the calculated radiative corrections with two external
fields $\Pi^{PP}(p^2)$ and $\Pi^{SS}(p^2)$. This is due to the fact
that after taking the Dirac trace in (\ref{eq: expansion}) the
interaction terms are from the start at most logarithmically divergent.
Recall that ambiguities emerge from the presence of quadratic
divergencies with external momentum dependence at the initial
stage (\ref{eq: ambiguidade}). Since the interaction Lagrangian
remains unchanged as compared to the one obtained in \cite{Mota},
we do not write it out here.

Now we display the renormalized propagators for the pions and scalar mesons
\begin{equation}
\label{eq: pionpropag}
   \Delta^{-1}_\pi = p^2 - \mu^2_{\pi}
   + 4N_cg^2_{\pi}F_{fin}(m, p^2; \lambda^2)
\end{equation}
\begin{equation}
\label{eq: sig propag}
   \Delta^{-1}_\sigma = p^2 - \mu^2_{\sigma}
   + 4N_cg^2_{\sigma}\Sigma_{fin}(m, p^2; \lambda^2)
\end{equation}
with the finite momentum dependent contributions
\begin{equation}
   F_{fin}(m,p^2;\lambda^2)=-p^2\left[Z_{0}(m^2,p^2;\lambda^2)
   -Z_{0}(m^2,0;\lambda^2)\right],
\end{equation}
\begin{eqnarray}
   \Sigma_{fin}(m,p^2;\lambda^2)&\!\!\!=\!\!\!&
   (4m^2-p^2)\left(Z_{0}(m^2,p^2;\lambda^2)
   -Z_{0}(m^2,0;\lambda^2)\right) \nonumber\\
   &\!\!\! -\!\!\!&4m^2 p^2 Z_{0}'(m^2,0;\lambda^2),
\end{eqnarray}
where use has been made of the normalization conditions
\begin{eqnarray}
\label{eq: normcond}
   && \Delta^{-1}_{\pi,\sigma}(0) = -\mu^2_{\pi,\sigma}\nonumber\\
   && \frac{d\Delta^{-1}_{\pi,\sigma}(p^2)}{dp^2}\Big |_{p^2 = 0} = 1 \mbox{.}
\end{eqnarray}
At the scale $\lambda^2=m^2$ one has $Z_{0}(m^2,0;m^2)=0$. We obtain
finally the physical pseudoscalar masses as zeros of these propagators
\begin{eqnarray}
\label{eq: physicalmasses}
   && m^2_{\pi} = \mu^2_\pi - 4N_cg^2_\pi
      F_{fin}(m,m^2_{\pi};\lambda^2) \nonumber\\
   && m^2_{\sigma} = \mu^2_{\sigma} - 4N_cg^2_\sigma
      \Sigma_{fin}(m,m^2_{\sigma};\lambda^2)
\end{eqnarray}
At the physical meson masses we obtain the following pseudoscalar
quark couplings
\begin{eqnarray}
\label{eq: couplingquarks}
   && g^{-2}_{\pi qq} = g^{-2}_{\pi} + 4N_c
      \frac{dF_{fin}(m,p^2;\lambda^2)}{dp^2}\Big |_{p^2 = m^2_{\pi}}\,
      , \nonumber \\
   && g^{-2}_{\sigma qq} = g^{-2}_{\sigma} + 4N_c
      \frac{d\Sigma_{fin}(m, p^2, \lambda^2)}{dp^2}\Big |_{p^2 =
      m^2_{\sigma}}\, .
\end{eqnarray}

\section{Coupling to external fields}
In order to evaluate the pion weak decay constant $f_\pi$, one
introduces the axial current as an external classical field. In the
present framework it appears in a generalized expression for the Dirac
operator, now containing the vector $V_{\mu}$ and axial vector field
$A_{\mu}$ \cite{Ebert:1986,OsipovHiller}; the relevant terms
for the calculation of $f_\pi$ are extracted after introducing the
covariant derivative for the pion
\begin{equation}
\label{eq: pionderiv}
   \partial_{\mu}\pi \rightarrow \nabla_{\mu}\pi_0 =
   \partial_{\mu}\pi_0 - 2\left(\sigma_0 +
   \frac{m}{g_0}\right) A_{\mu} - i[V_{\mu}, \pi_0] ,
\end{equation}
in $Y$, eq.(\ref{Y}) and in the kinetic terms
$$
  \frac{f^{2}_0}{2}\,\mbox{Tr}\,{\nabla_{\mu}\pi_0\nabla^{\mu}\pi_0}
  \rightarrow - f^2_{0}\frac{m}{g_0}A^{\mu}\partial_{\mu}\pi_0.
$$

The expression for the pion decay constant is related to the coefficient of
$p^{\mu}A_{\mu}\pi^{ph}$ where the physical pion field
 $\pi^{ph}$ is introduced through  $$\pi_0 =
\pi^{ph}\frac{g_{\pi qq}}{g_0},$$ and reads
\begin{equation}
\label {eq: fpi}
   f_{\pi} = \frac{m g_{\pi qq}}{g_0^2}\left[f^2_0
   -4N_cg^2_0 \left(I_{log}(\lambda^2)
   + Z_{0}(m^2,m^2_{\pi};\lambda^2)\right)\right].
\end{equation}
Note that the contribution of the radiative correction to $f_{\pi}$
does not involve the ambiguity $\alpha$, as opposed to the other
radiative corrections with two external fields $\Pi^{PP}(p^2)$ and
$\Pi^{SS}(p^2)$. This is again due to the fact that the Dirac trace
renders the expression for the pseudoscalar-axial bubble
$\Pi_{\mu}^{PA}$ logarithmically divergent
\begin{equation}
\label{PAbubble}
   \Pi_{\mu}^{PA}= -i\elemint \frac{4mp_\mu}{(k^2 - m^2)[(k-p)^2-m^2]}
\end{equation}

Using eq.(\ref{eq: physicalmasses})
\begin{equation}
   1-\frac{\mu^2_\pi}{m_\pi^2}=4N_c g_\pi^2
   \left[Z_0(m^2,m_\pi^2;\lambda^2)-Z_0(m^2,0;\lambda^2)\right]
\end{equation}
together with eqs.(\ref{eq: constren2}) and (\ref{eq: gpi}) in
eq.(\ref{eq: fpi}) one gets
\begin{equation}
\label{fpif}
   f_\pi = m\frac{g_{\pi qq}\mu^2_{\pi}}{g^2_\pi m^2_{\pi}} + \alpha'
\end{equation}
where $\alpha'$ is related to $\alpha$ by
$$
   \alpha' = -4N_c m g_{\pi qq}\alpha .
$$
It is now obvious that $\alpha'$ must be zero, given the
Goldberger-Treiman relation, valid in the chiral limit
\begin{equation}
   g_\pi f_{\pi} = m \mbox{.}
\end{equation}
This relation is obtained from (\ref{fpif}) with $\alpha'=0$, since in
the chiral limit $m_{\pi} = \mu_{\pi}$ as can be seen from
(\ref{eq: physicalmasses}) and $g_{\pi qq}=g_\pi$ from
(\ref{eq: couplingquarks}).

\section{Conclusion}
In this contribution we extended the work presented in \cite{Mota}
in a way as to leave open until the end of the calculations all
arbitrary finite constants related with differences of divergent
integrals of the same degree of divergence. This is possible using
the implicit regularization technique, in which the divergent content
of any amplitude is rendered strictly independent of the external
(physical) momenta. It turns out that renormalization absorbs some of
the arbitrary constants, while chiral symmetry fixes the arbitrary
parameter appearing in the pion decay constant to be zero.
\vspace{0.5cm}

{\bf Acknowledgements}
\vspace{0.5cm}

\noindent
This work has been supported by grants provided by CAPEs - Brazil and
Funda\c c\~ao para a Ci\^encia e a Tecnologia, Portugal,
POCTI/FNU/50336/2003 and PRAXIS XXI/BCC/4301/94.
This research is part of the EU integrated infrastructure initiative
HadronPhysics project under contract No.RII3-CT-2004-506078.
A. A. Osipov also gratefully acknowledges the
Funda\ca o Calouste Gulbenkian for financial support.

\end{document}